\documentclass[aps,superscriptadress,preprint]{revtex4}
\usepackage{graphicx}

\begin{document}

\title{Extended quark mass density- and temperature- dependent model and the deconfinement phase transition}
\author{Wei Liang Qian\footnote{wlqian@fudan.edu.cn}}
\affiliation{Department of Physics, Fudan University, Shanghai 200433, P.R.China}
\author{Ru-Keng Su\footnote{rksu@fudan.ac.cn}}
\affiliation{China Center of Advanced Science and Technology (World Laboratory), P. O.
Box 8730, Beijing 100080, P.R.China}
\affiliation{Department of Physics, Fudan University, Shanghai 200433, P.R.China}

\begin{abstract}
An extended quark mass density- and temperature- dependent model which
includes the couplings between quarks and the $\sigma$-mesons, $\omega$-mesons
is suggested.
The MIT bag boundary constrain has been given up and the interactions between
quarks and mesons are extended to the whole free space.
We show that the present model is successful to describe both the
saturation properties and the deconfinement phase transition of nuclear matter.
When the effective nucleon masses vanish and the bag radius tends to infinite,
the quark deconfinement phase transition takes place. The corresponding QGP phase diagram
is addressed.
\end{abstract}

\maketitle

\newpage

\section{Introduction}
Because of the non-perturbative properties of QCD at low energy
regions, it is very difficult to study nuclear system by using QCD
as a fundamental theory directly. Various phenomenological models
based either on hadron degree of freedom or quark degree of freedom
have been suggested. The quark-meson coupling (QMC) model, first
proposed by Guichon\cite{g88}, which describes nuclear matter as a
collection of non-overlapping MIT bags,
scalar $\sigma$ meson and vector $\omega$ meson is one of such successful
candidates. The quarks inside the MIT bag couple with the scalar
$\sigma$ meson and vector $\omega$ meson self-consistently. By means
of this model and the mean field approximation, many dynamical and
thermal properties of nucleon systems and hyperon systems have been
studied both at zero temperature and finite
temperature\cite{s94,j96,s95,za00,w99,s03,p03}.

Although QMC model is successful for describing the physical properties of nuclear
system, many shortcomings arise when one use this model to discuss the
quark deconfinement. The first difficulty comes from that this is a permanent quark
confinement model. The nucleon corresponds to a MIT bag and the mechanism of quark confinement in MIT
bag model being that the normal flow of quark current is zero at the bag surface.
This boundary condition can not be changed by temperature and density.
The second difficulty arises from the many-body calculations. If we hope to
do the nuclear many-body calculations beyond mean field approximation by
quantum field theory, it is essential to find the propagators of quark, $\sigma$ meson
and $\omega$ meson respectively. But the constrain of MIT bag boundary condition
and the interaction between quarks and mesons limited within the bag regions
present obstacles to get the corresponding propagators in free space.

On the other hand, another effective model, namely, the quark mass
density- depedent (QMDD) model, was suggested by Fowler, Raha and Weiner\cite{frw81}.
According to this model, the masses of u, d quarks and strange s quark (togather with the corresponding
anti-quarks) are given by
\begin{eqnarray}
m_q &=& \frac{B}{n_Q} \ \ \ \ \ (q = u, d, \bar{u}, \bar{d})   \\
m_{s, \bar{s}} &=& m_{s0} + \frac{B}{n_Q}
\end{eqnarray}
where $B$ is the vacuum energy density inside the bag and $n_Q$ is the quark
number density. The basic hypothesis Eqs.(1) and (2) corresponds to a quark
confinement mechanism because if quark goes to infinite, the volume of the system
tends to infinite, $n_Q$ approaches zero and $m_q$ approaches to infinite.
The infinite quark mass prevents the quark from going to infinite\cite{bl95,zs01}.
This confinement mechanism is very similar to that of the MIT bag model.
As was shown in ref.\cite{bl95}, the properties of strange matter in
QMDD model are nearly the same as those obtained by the MIT bag model.
Although the confinement mechanism is similar, the advantage of QMDD model
is that it does not need to introduce a boundary condition to confine quark
as that of the MIT bag model.

As was pointed by refs.\cite{zs02,zs031}, if we use QMDD model to investigate
the thermodynamical properties of nuclear system at finite temperature, many
difficulties will emerge. For example, it cannot reproduce a correct
lattice QCD phase diagram qualitatively because the masses of quarks,
and then the temperature tends to infinite when $n_B\rightarrow 0$\cite{zs02}.
The reason is that the confinement mechanism of QMDD model is still permanent.
To overcome this difficulty, in a series of our pervious papers\cite{zs01,zs02,zs031,zs032,zs04,w05},
we suggested a quark mass density- and temperature- dependent (QMDTD) model.
Instead of the permanent MIT quark confinement mechanism, we employed the
Friedberg-Lee (FL) soliton bag model. Since the confinement mechanism of
FL model comes from the interaction between quarks and a nontopological scalar
soliton field, and the spontaneously broken symmetry of the scalar field will
be restored at a finite temperature, the soliton solution will disappear and
the quark will deconfine at the critical temperature. In this nonpermanent quark confinement
model, the vacuum energy density $B$ equals the different value between the
perturbative vacuum and physical vacuum, and it depends on temperature. Instead of Eqs.(1) and (2),
we introduced\cite{zs01,zs02,zs031,zs032,zs04,w05}
\begin{eqnarray}
B(T) &=& B_0\left[ 1-\left( \frac T{T_c}\right)^2\right] \ \ \ \ 0 \leq T \leq T_c   \\
B(T) &=& 0\ \ \ \ T > T_c
\end{eqnarray}
and
\begin{eqnarray}
m_q &=& \frac{B_0}{n_Q}\left[ 1-\left( \frac T{T_c}\right)^2\right] \ \ (q = u, d, \bar{u}, \bar{d})\ \ \ \ 0 \leq T \leq T_c\\
m_q &=& 0\ \ \ \ T > T_c\\
m_{s, \bar{s}} &=& m_{s0} + \frac{B_0}{n_Q}\left[ 1-\left( \frac T{T_c}\right)^2\right]\ \ \ \ 0 \leq T \leq T_c\\
m_{s, \bar{s}} &=& m_{s0}\ \ \ \ T > T_c
\end{eqnarray}
in QMDTD model. The masses of quarks not only depend on density but also on temperature.
Obviously, the QMDTD model reduces to QMDD model at zero temperature.
By means of the QMDTD model, the physical properties and the stability of strangelets\cite{zs031},
the dibaryon system\cite{zs04} and the strange star\cite{s05,g03} at finite temperature have been
studied and the results are fine.

Although the masses of quarks reflect the confinement characteristic in the QMDTD model,
it is still an ideal quark gas model. If we hope to investigate the
saturation properties and the deconfinement phase transition of nuclear matter,
the quark-quark interaction must be taken into consideration. In ref.\cite{w05}, a quark and non-linear
scalar field coupling is introduced to improved QMDD model at zero temperature. We proved that
many physical properties given by FL soliton bag model can be mimicked by
the improved QMDD model.

This paper evolves from an attempt to extend our study to finite
temperature. We will introduce the scalar $\sigma$ meson, vector
$\omega$ meson and the couplings between quarks (u, d) and $\sigma$
meson, $\omega$ meson in the extended quark mass density- and
temperature- dependent (EQMDTD) model. Our EQMDTD model is similar
to that of Walecka model and QMC model. The basic differences
between the EQMDTD model and the Walecka model are: (1)we replace
the nucleon in Walecka model by quark in EQMDTD model; (2)rather
than the structureless point-like nucleon in Walecka model, the
nucleon corresponds to a "cluster", say, bag which consists three
quarks with temperature- and density- dependent masses given by
Eqs(5) and (6). Besides, the differences between the EQMDTD model
and the QMC model are: (1)instead of the MIT bag in QMC model, we
incorporate quarks with density- and temperature- dependent masses
in EQMDTD model; (2)in place of the quark-$\sigma$ quark-$\omega$
interactions restricted within the bag region in QMC model, the
quark-$\sigma$, quark-$\omega$ interactions are spreaded to the
whole free space, because the constraint of MIT bag boundary
condition is discarded. Moreover, since our model are based on the
FL model, the quark deconfinement phase transition can take place.
we will address the behavior of the saturation and the deconfinement
phase transition for EQMDTD model in this paper.

The organization of this paper is as follows. In the next section, we will
give a brief description of EQMDTD model and the main formulae. Our numerical results
will be presented in Section III. The last section is devoted to discussion and conclusion.

\section{The EQMDTD Model}

The Lagrangian density of the EQMDTD model is
\begin{eqnarray}
{\cal L}  = \bar{\psi} \left[ i\gamma^{\mu} \partial_{\mu}-m_q
        +(g_{\sigma}^q \sigma - g_{\omega}^q \gamma^{\mu} \omega_{\mu})
          \right] \psi
    + \frac{1}{2}\partial_{\mu}\sigma\partial^{\mu}\sigma
        - \frac{1}{2}m_{\sigma}^2 \sigma^2
        - \frac{1}{4}F_{\mu\nu}F^{\mu\nu}
        + \frac{1}{2}m_{\omega}^2 \omega_{\mu}\omega^{\mu}     \label{La}
\end{eqnarray}
where the quark mass $m_q$ is given by Eqs.(5) and (6), $m_\sigma$, $m_\omega$ are the masses
of $\sigma$ and $\omega$ mesons, $F_{\mu\nu}=\partial_{\mu}\omega_{\nu}-\partial_{\nu}\omega_{\mu}$,
$g_{\sigma}^q$ and $g_{\omega}^q$ are the couplings
between quark-$\sigma$ meson and quark-$\omega$ meson respectively. We neglect the s quark in
the following discussion.

The equation of motion for quark field is
\begin{equation} \label{EL}
  \left[ \gamma^{\mu} \left(i\partial_{\mu}+ g_{\omega}^q \omega_{\mu} \right)
        - \left( m_q - g_{\sigma}^q \sigma \right)
          \right] \psi = 0
\end{equation}
Under mean field approximation, the effective quark mass $m_q^*$ is given by
\begin{equation}
m_q^* = m_q - g_{\sigma}^q \bar{\sigma}
\end{equation}
In nuclear matter, three quarks constitute a bag, say nucleon,
and the effective nucleon mass is obtained from the bag energy and reads
\begin{equation}
M_N^* = E_{bag} = \sum_q E_q = \frac{4}{3}\gamma \pi R^3 \sum _{q = u,d} \sum _{k}  \varepsilon _q(k)(f_q(k)+\bar{f_q}(k))
\end{equation}
where $\varepsilon _q(k)=\sqrt{m_q^{*2}+k^2}$ is the single
particle energy, $\gamma$ is the degeneracy,
$f_q(k)$ and $\bar{f_q}(k)$ are the Fermi distributions of quark and anti-quark respectively
\begin{eqnarray}
f_q(k) &=& \frac{1}{e^{\beta(\varepsilon _q(k)-\mu_q)}+1} \\
\bar{f_q}(k) &=& \frac{1}{e^{\beta(\varepsilon _q(k)+\mu_q)}+1}
\end{eqnarray}
The chemical potential $\mu_q$ of quark is given by
\begin{eqnarray}
n_Q &=& \gamma \sum_{q=u,d} \sum_k (f_q(k)-\bar{f_q}(k))  \\
3 &=& \frac{4}{3} \pi R^3 n_Q
\end{eqnarray}
The bag radius $R$ is determined by the equilibrium condition for the nucleon bag
\begin{equation}
\frac{\delta M_N^*}{\delta R} = 0
\end{equation}

In nuclear matter, the total energy density is given by\cite{s94}
\begin{equation}
{\cal{E}}_{matter} = \sum _{i=N,P} \frac{2}{(2\pi)^3} \int d^{3}k\sqrt{M_i^{*2}+k^2}\left( n_i(k)+\bar{n}_i(k) \right)
+ \frac{g_\omega^2}{2m_\omega^2}\rho_B^2
+ \frac{1}{2} m_\sigma^2\bar{\sigma}^2  \label{em}
\end{equation}
where $\rho_B$ is the density of nuclear matter
\begin{equation}
\rho_B = \sum _i \frac{2}{(2\pi)^3} \int d^{3}k \left( n_i(k)-\bar{n}_i(k) \right)
\end{equation}
$n_i(k)$ and $\bar{n}_i(k)$ are the Fermi distributions of nucleon
and anti-nucleon respectively,
\begin{eqnarray}
n_i(k) &=& \frac{1}{e^{\beta(\sqrt{M_N^{*2}+k^2}-\mu_i)}+1} \\
\bar{n}_i(k) &=& \frac{1}{e^{\beta(\sqrt{M_N^{*2}+k^2}+\mu_i)}+1}
\end{eqnarray}
$\mu_i$ is the chemical potential of the ith nucleon. In Eq.(\ref{em}), $g_\omega$
is the coupling between nucleon and $\omega$ meson and it satisfies
$g_\omega = 3g_\omega^q$. As that of QMC model\cite{s94,j96,s95,za00,w99,s03,p03}
the scalar mean field of ${\sigma}$ meson is determined by self-consistency condition
\begin{eqnarray}
\left.\frac{\delta E_{matter}}{\delta \sigma}\right|_{\sigma=\bar{\sigma}} = 0
\end{eqnarray}
which yields
\begin{equation}
\bar{\sigma} = -\frac{2}{m_\sigma^2(2\pi)^3} \sum_{i=N,P} \int d^{3}k
\frac{M_i^*}{\sqrt{M_i^{*2}+k^2}}
\left( \frac{\partial{M_i^*}}{\partial \bar{\sigma}} \right)_{bag, T} \left( n_i(k)+\bar{n}_i(k) \right) \label{scc}
\end{equation}
Eqs.(5),(9)-(20) form a complete set of equations and we can solve them numerically. Our numerical results
will be shown in the next section.

\section{Results}
Before numerical calculation, let us discuss the effective quark mass $m_q^*$ carefully because this
quantity affects our results directly. To take the medium effects into account, Jin and Jinnings
suggested a modified QMC model\cite{j96} in which the bag parameter $B$ depends on density. In EQMDTD model,
based on FL model, the bag parameter $B$ depends on temperature (Eq.(3)), and the quark
mass $m_q$ depends on quark density $n_Q$. As was shown in section I, this quark density dependence
of $m_q$ corresponds to a quark confinement mechanism only. The medium effect has not
yet been considered. To exhibit the medium effects, instead of $B = B(n_Q)$
in the modified QMC model, we introduce an
ansatz that the effective coupling of quarks and $\sigma$ meson $g_{\sigma}^q$ is a function of
quark density $n_Q$, which can be expanded as
\begin{equation}
g_{\sigma}^q = g_{\sigma}^{q(0)}/n_Q + g_{\sigma}^{q(1)}/n_Q^2 + g_{\sigma}^{q(2)}/n_Q^3     \label{gcv}
\end{equation}
Eq.(9) becomes
\begin{equation}
m_q^* = {\frac{B_0\left[1 - \left(T/{T_c}\right)^2\right]-(g_{\sigma}^{q(0)} + g_{\sigma}^{q(1)}/n_Q + g_{\sigma}^{q(2)}/n_Q^2)\bar{\sigma}}{n_Q}}
\end{equation}
In fact, the density- dependent couplings of $NN\pi$, $NN\rho$, $NN\sigma$ have been
employed by many authors to discuss many different problems\cite{q93,s93,q00,m04,s01}.
Eq.(\ref{gcv}) is an extension to quark level only. The values of the adjusted parameters
$g_{\sigma}^{q(0)}$, $g_{\sigma}^{q(1)}$ and $g_{\sigma}^{q(2)}$ will be determined below.

Now we discuss the parameters in EQMDTD model. Firstly, we choose
$m_\omega = 783$ MeV, $m_\sigma = 509$ MeV as that of
refs.\cite{zl00,fst}; $T_c = 170$ MeV as the deconfinement
temperature at zero baryon density. To fix the nucleon mass $M_N =
939$ MeV, we take $B_0 = 173$ MeVfm$^{-3}$. Beside these parameters,
there are still four parameters, namely, $g_{\omega}^q$,
$g_{\sigma}^{q(0)}$, $g_{\sigma}^{q(1)}$ and $g_{\sigma}^{q(2)}$
needed to be fixed. Obviously, the behavior at the saturation point
and the deconfinement phase transition must be explained if EQMDTD
model is successful. We see below that the quark deconfinement phase
transition will take place at the point $M_N^*\rightarrow 0$ and bag
radius $R\rightarrow \infty$. Therefore, we fix parameters
$g_{\omega}^q$, $g_{\sigma}^{q(0)}$, $g_{\sigma}^{q(1)}$ and
$g_{\sigma}^{q(2)}$ by the condition that at zero temperature,
binding energy $E = - 15$ MeV, the ratio of the effective nucleon
mass to the nucleon mass $M_N^*/M_N = 0.6$ at the saturation density
$\rho_0 = 0.15$ fm$^{-3}$, and the deconfinment phase transition
takes place at $\rho_B = 10\rho_0$. We find
\begin{eqnarray}
g_{\sigma}^{q(0)} = 5.93 fm^{-3}, \ \ \ \ g_{\sigma}^{q(1)} = - 0.747 fm^{-6},\ \ \ \
g_{\sigma}^{q(2)} = 0.035 fm^{-9}, \ \ \ \  g_{\omega}^q = 4.23 \nonumber
\end{eqnarray}

Now we are in a position to do numerical calculations and the
results are shwon in Fig.1a-Fig.9b. In Fig.1a, we omit the
contribution of $\sigma$-meson and $\omega$-meson and depict the bag
energy as a function of bag radius at different temperatures $T = 0,
30, 60, 100, 150, 170$ MeV respectively. In this specific case, the
EQMDTD model reduces to the original QMDTD model without quark-meson
interaction. Fig.1b is the same diagram as that of Fig.1a but that
contribution of meson fields has been taken into consideration. We
fix $\bar{\sigma} = 20$ MeV. It is found from Fig.1a and Fig.1b that
the radius of the bag given by the minimum of the curve increases
with increasing temperature. When $T < 100 MeV$, the bag radius
varies merely slightly with increasing temperature. This result is
similar to that of the QMC model or the FL model. However, when the
temperature approaches to the critical temperature $t_c$, the bag
radius increases significantly.

Noting that the $\bar{\sigma}$ connects to the baryon density
$\rho_B$ by Eq.(23) directly, instead of $\rho_B$, we employ
$\bar{\sigma}$ to show our results first. The curves of effective
nuclear mass $M^*$ vs. the mean field $\bar{\sigma}$ at different
temperatures $T = 0, 30, 60, 100, 130, 160$ MeV are shown in Fig.2.
We see that the effective nucleon mass decreases when $\bar{\sigma}$
increases. In particular, we find a very interesting phenomenon:
when $\bar{\sigma}$ approaches to a fixed point $\bar{\sigma}_c$,
$M^*$ drops to zero rapidly. This fixed point $\bar{\sigma}_c$
decreases as the temperature increases. The result $M^* = 0$ means
that the nucleon is dissolved and the quark deconfinement phase
transition takes place. To show this phase transition more clearly,
we draw the curves $R$ vs. $\bar{\sigma}$ at different temperatures
$T = 0, 30, 60, 100, 130, 160$ MeV in Fig.3. We see that at the same
critical point $\bar{\sigma}_c$ shown by Fig.2, the bag radius $R$
tends to infinite rapidly. This result confirms that this is indeed
a critical point of deconfinement phase transition, because at this
case quarks occupy the whole system. Similarly, for fixed
$\bar{\sigma} = 0, 20, 40, 60, 70, 75$ MeV, we show $M^*$ vs. $T$
curves and $R$ vs. $T$ curves in Fig.4 and Fig.5 respectively. We
see that for every curve, $M^*$ drops to zero and $R$ tends to
infinite at a critical temperature $t_c$. The critical temperature
$t_c$ increases when $\bar{\sigma}_c$ decreases. There results
exhibit that the quark deconfinement phase transition can be
described by EQMDTD model naturally.

Now let us show the connection between $\bar{\sigma}_c$ and the baryon density
$\rho_B$. Employing Eqs. (19) and (23), one can obtain the relation of
$\bar{\sigma}$ and $\rho_B$. The result is shown in Fig.6. For a fixed
temperature, $\bar{\sigma}$ is a monotonical function of $\rho_B$. It means
that $\bar{\sigma}$ can play the same role as that of $\rho_B$. For example,
$M^*$ vs. $\bar{\sigma}$ curve has the same behavior as that of $M^*$ vs. $\rho_B$
curve. The curve in Fig.7 shows that when $\bar{\sigma}_c$ increases,
the critical temperature $t_c$ decreases. It corresponds to the result
that when $\rho_B^c$ increases, $t_c$ decreases for the quark deconfinement phase
transition. This is of course very reasonable. In fact, Fig.7 is the
deconfinement phase diagram for EQMDTD model. It is very similar to that
given by lattice calculation.

Now we turn to address the saturation properties of nuclear matter at low
temperature and low density by using the EQMDTD model. Our results are shown
in Fig.8, Fig.9a and Fig.9b. The curves of ${\cal{E}}_{matter}/\rho_B - M_N$ vs. $\rho_B$
for different temperatures are shown in Fig.8 where the curve at $T = 0$ MeV refers
to the saturation curve and the corresponding minimum is the saturation point.
The changes of the effective nucleon mass $M^*$ to $\rho_B$ are shown in Fig.9a
and Fig.9b where the curves in Fig.9a limit to the low density region ($0$ fm$^{-3}\le \rho_B \le 0.20$ fm$^{-3}$),
and curves in Fig.9b cover a large density region. We find that $M^*$
decreases when density and/or temperature increases. This result
is similar to the of QMC model[47] and Walecka model. At the saturation point,
$M^*/M = 0.6$, as indicated.

\section{Summary}
In summary, we present an extended quark mass density- and
temperature- dependent model which is motivated by the QMDTD model
and the quark meson coupling model. The $\sigma$ and $\omega$ fields
are assumed to couple with the $u$, $d$ quarks. The MIT boundary
constrain has been given up and the non-permanent quark confinement
mechanism is mimicked by the quark mass density- and temperature-
dependence. It is shown that under mean field approximation this
model provides a reasonable description of the saturation properties
for nuclear matter in the region of low temperature and density.
While at sufficient high temperature and/or density, the
deconfinement phase transition takes place automatically. The phase
transition is characterized by that the bag radius tends to infinite
and the effective baryon mass $M^*$ vanishes simultaneously. The
deconfinement phase diagram which has the same behavior as that of
lattice QCD calculation is addressed. We emphasize that in our
calculation the deconfinement process is treated consistently within
one unified model. This is different from the usual investigation
which employs two separate models to describe respectively the
hadronic phase and quark gluon plasma phase, and use the Gibbs
conditions to determine the phase diagram[26,27].

\section*{Acknowledgments}
This work is supported in part by National Natural Science
Foundation of China under No.10375013, 10405008, 10247001,
10235030, National Basic Research Program of China 2003CB716300,
and the Foundation of Education Ministry of China 20030246005.

\begin{figure}[tbp]
\includegraphics[scale=0.8]{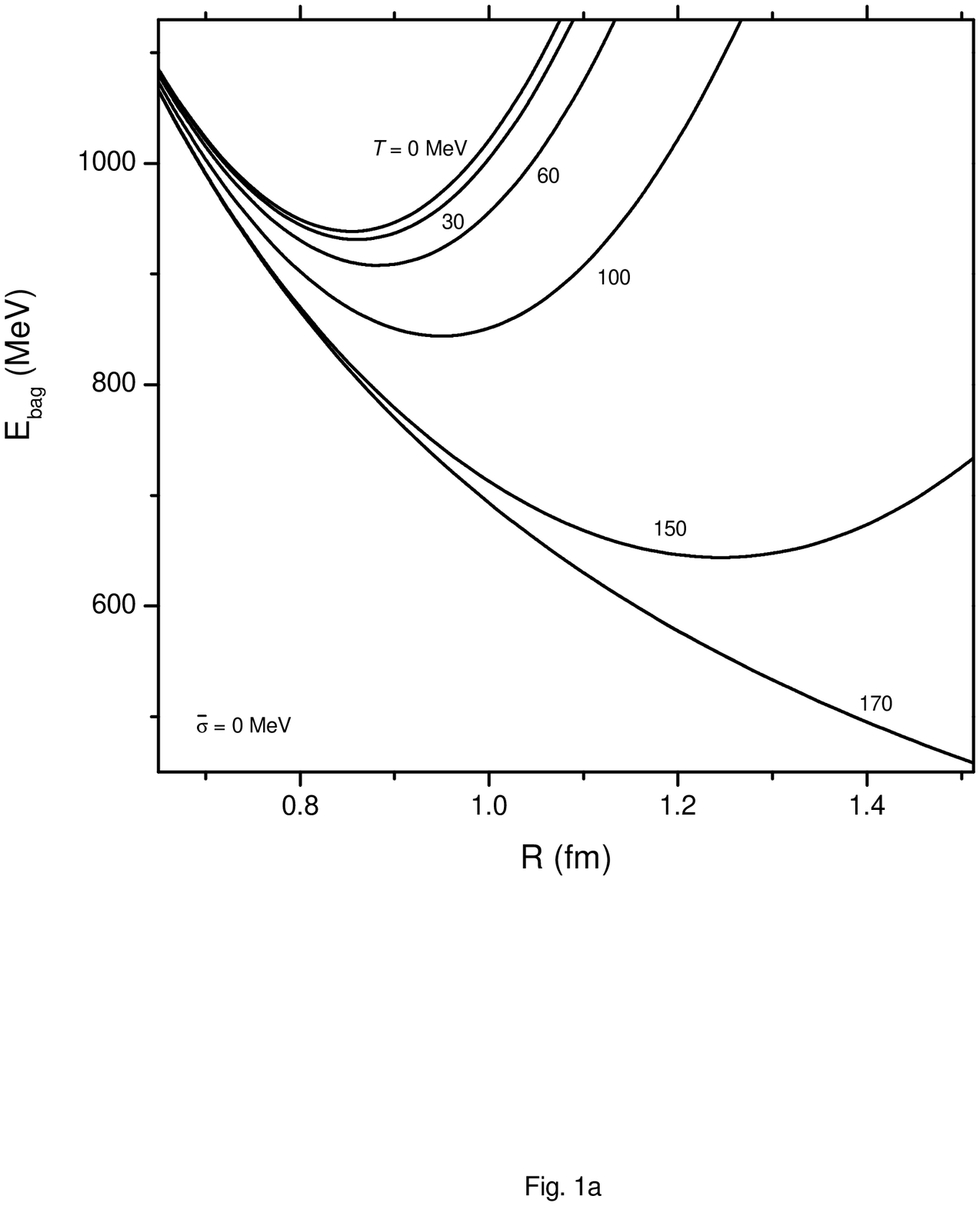}
\caption{The bag energy as a function of bag radius without meson at different temperatures
$T = 0, 30, 60, 100, 150$ and $168$ MeV, with $\bar{\sigma} = 0$ MeV, the radius for nucleon is determined by the minimum of
bag energy.}
\label{fig1a}
\end{figure}

\begin{figure}[tbp]
\includegraphics[scale=0.8]{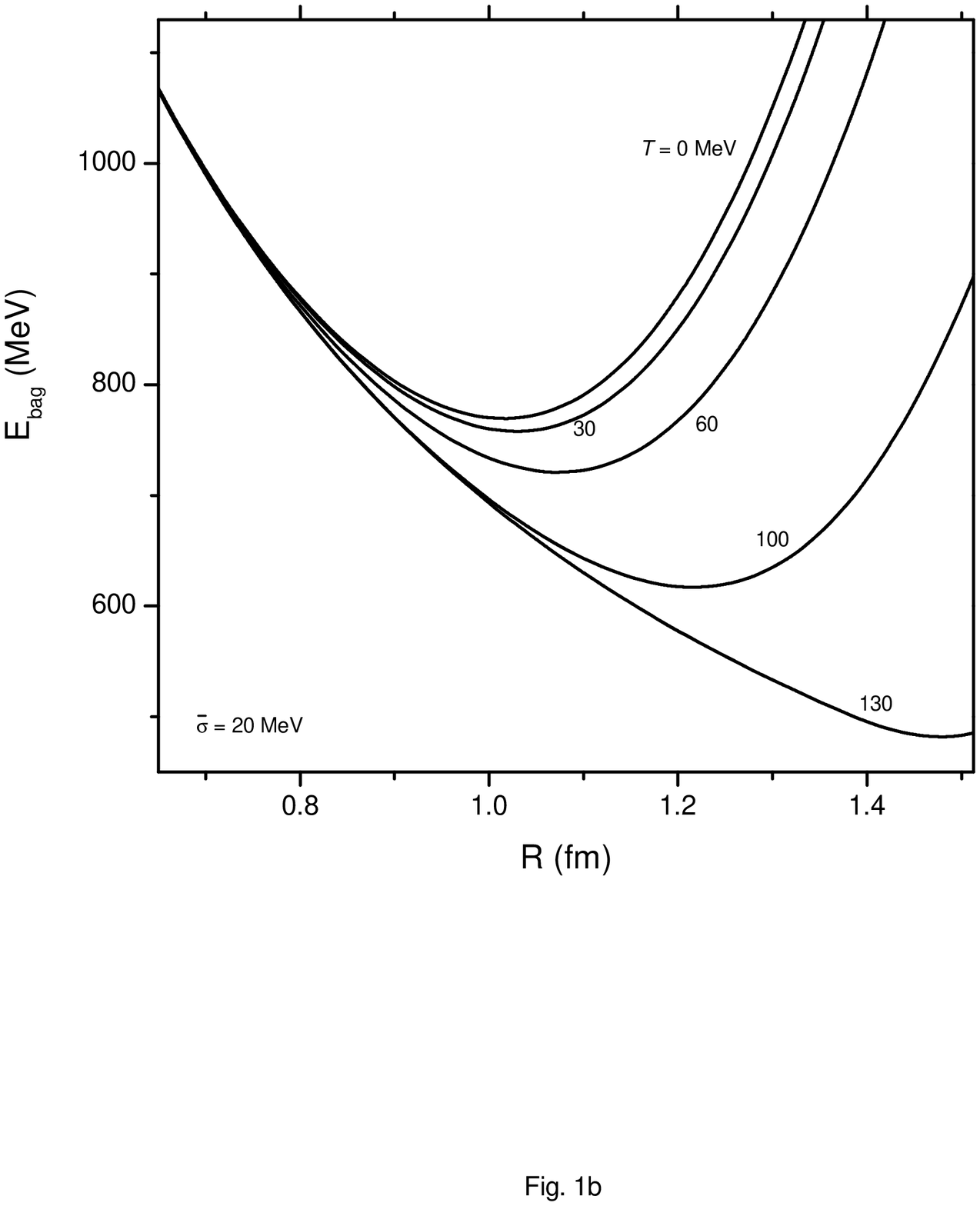}
\caption{The bag energy as a function of bag radius without meson at different temperatures
$T = 0, 30, 60, 100, 150$ and $170$ MeV, with $\bar{\sigma} = 20$ MeV.}
\label{fig1b}
\end{figure}

\begin{figure}[tbp]
\includegraphics[scale=0.8]{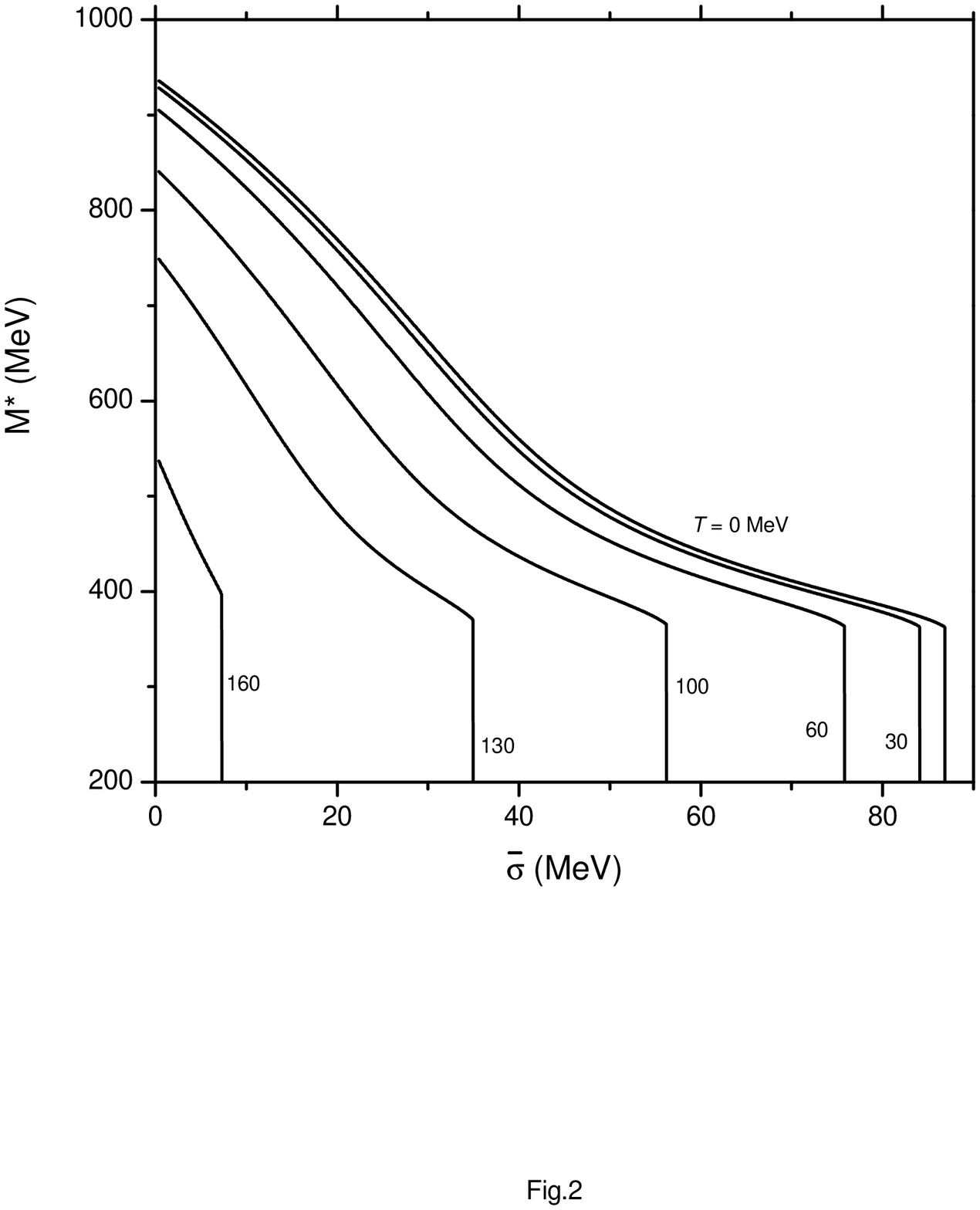}
\label{fig2}
\caption{Effective nucleon masses vs. the $\bar{\sigma}$ at different temperatures
$T = 0, 30, 60, 100, 130$ and $160$ MeV respectively}
\end{figure}

\begin{figure}[tbp]
\includegraphics[scale=0.8]{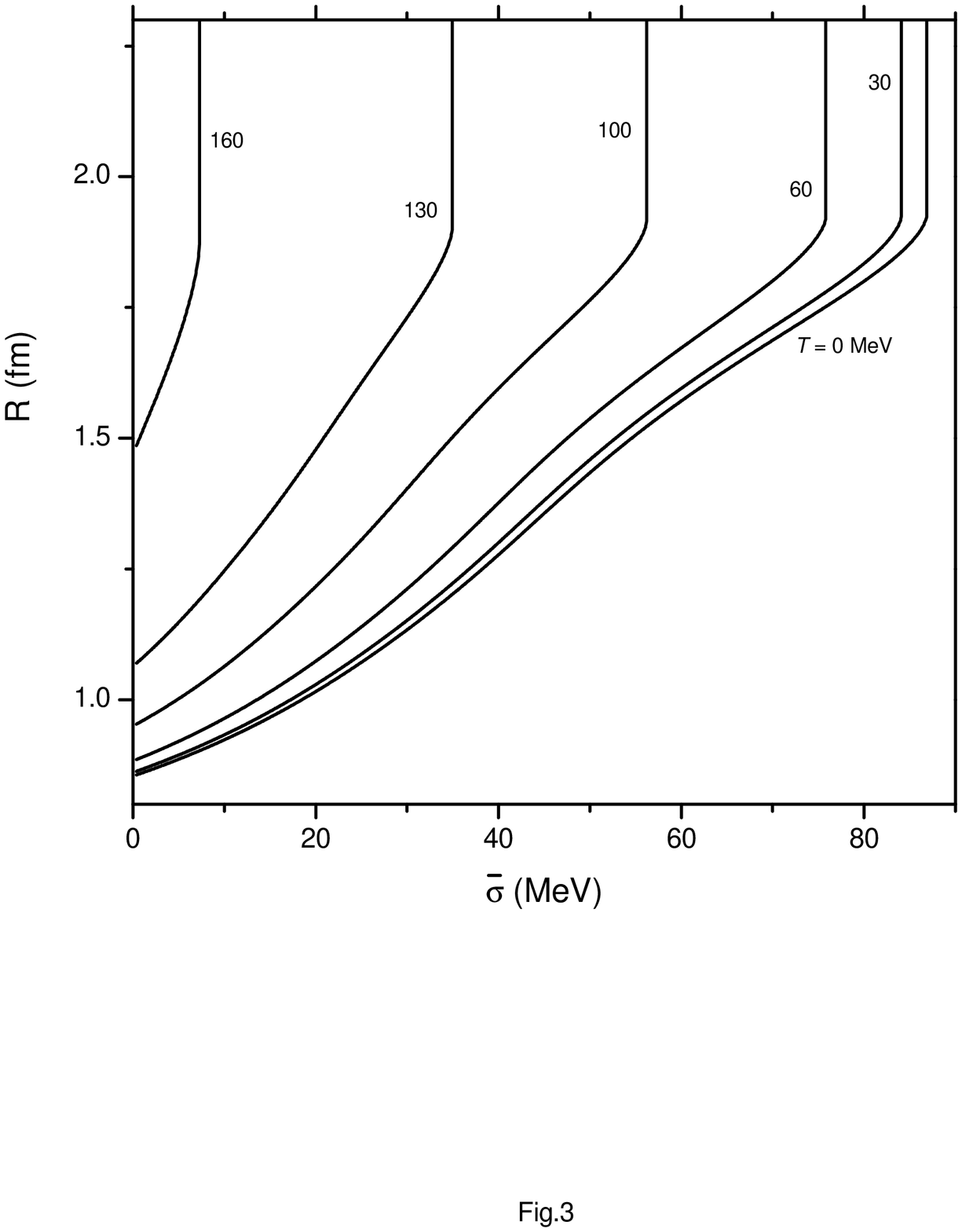}
\label{fig3}
\caption{Nucleon radius vs. the $\bar{\sigma}$ at different temperatures
$T = 0, 30, 60, 100, 130$ and $160$ MeV respectively}
\end{figure}

\begin{figure}[tbp]
\includegraphics[scale=0.8]{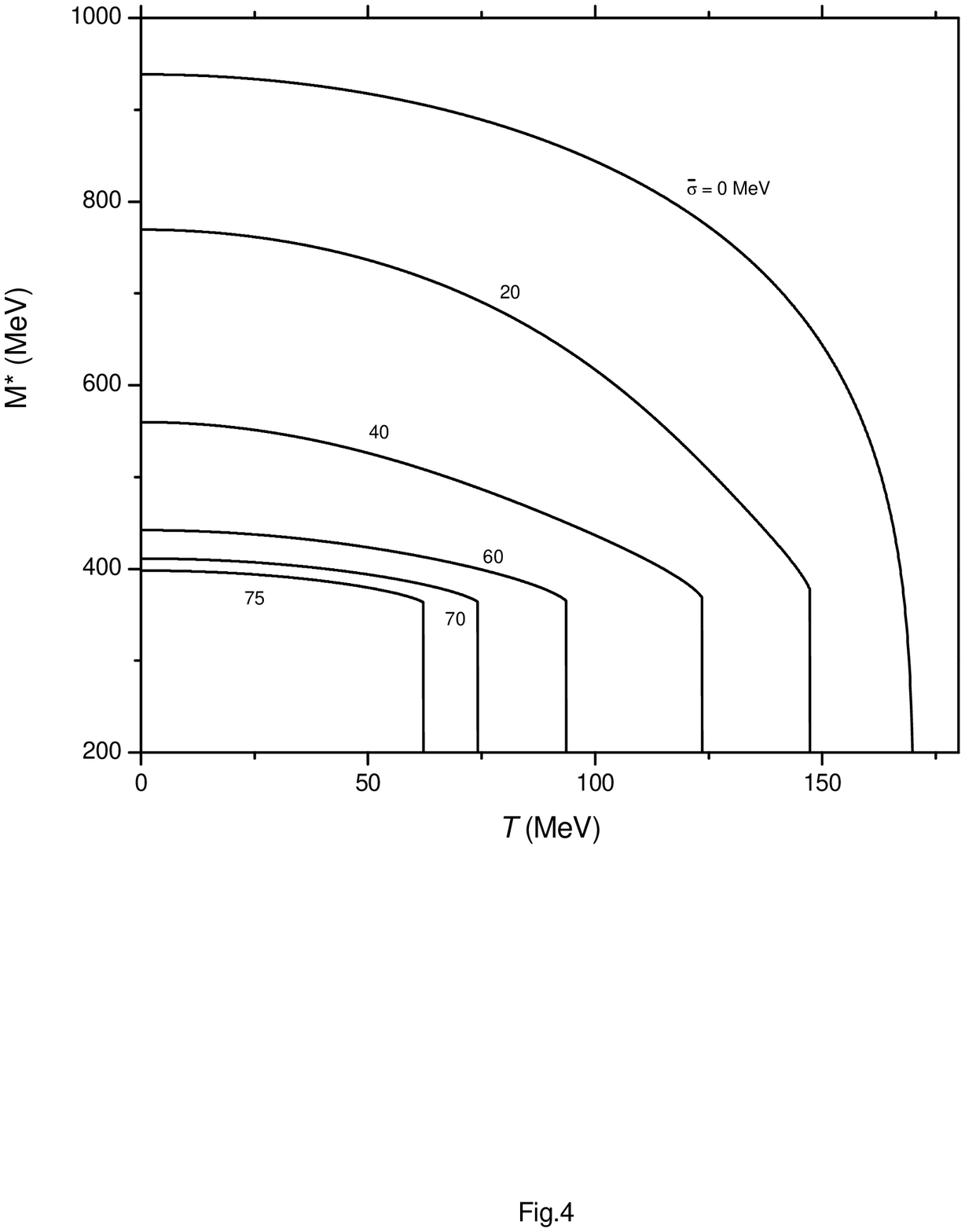}
\label{fig4}
\caption{Effective nucleon masses vs. the temperatures at different values
$\bar{\sigma} = 0, 20, 40, 60, 70$ and $75$ MeV respectively}
\end{figure}

\begin{figure}[tbp]
\includegraphics[scale=0.8]{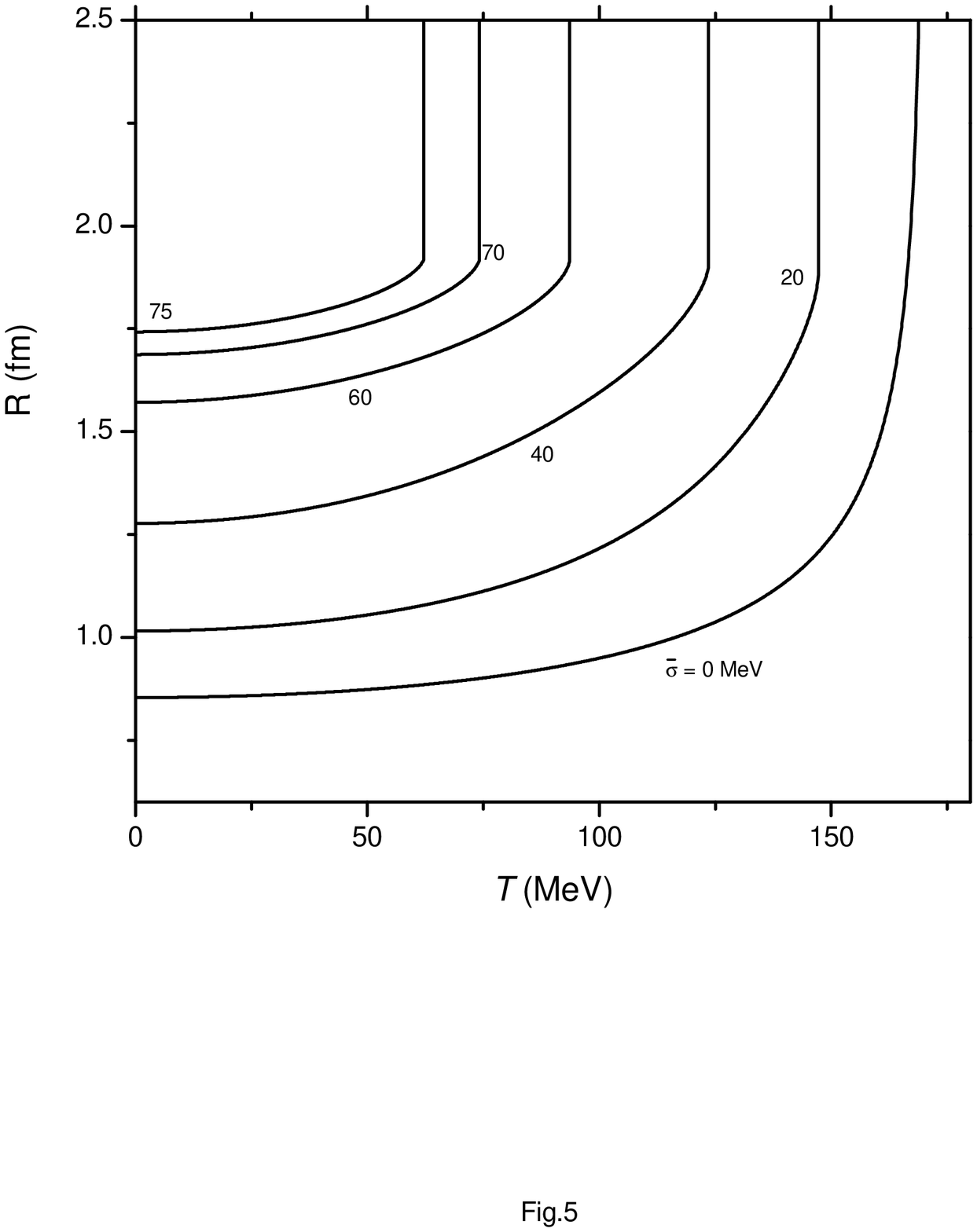}
\label{fig5}
\caption{Nucleon radius vs. the temperatures at different values
$\bar{\sigma} = 0, 20, 40, 60, 70$ and $75$ MeV respectively}
\end{figure}

\begin{figure}[tbp]
\includegraphics[scale=0.8]{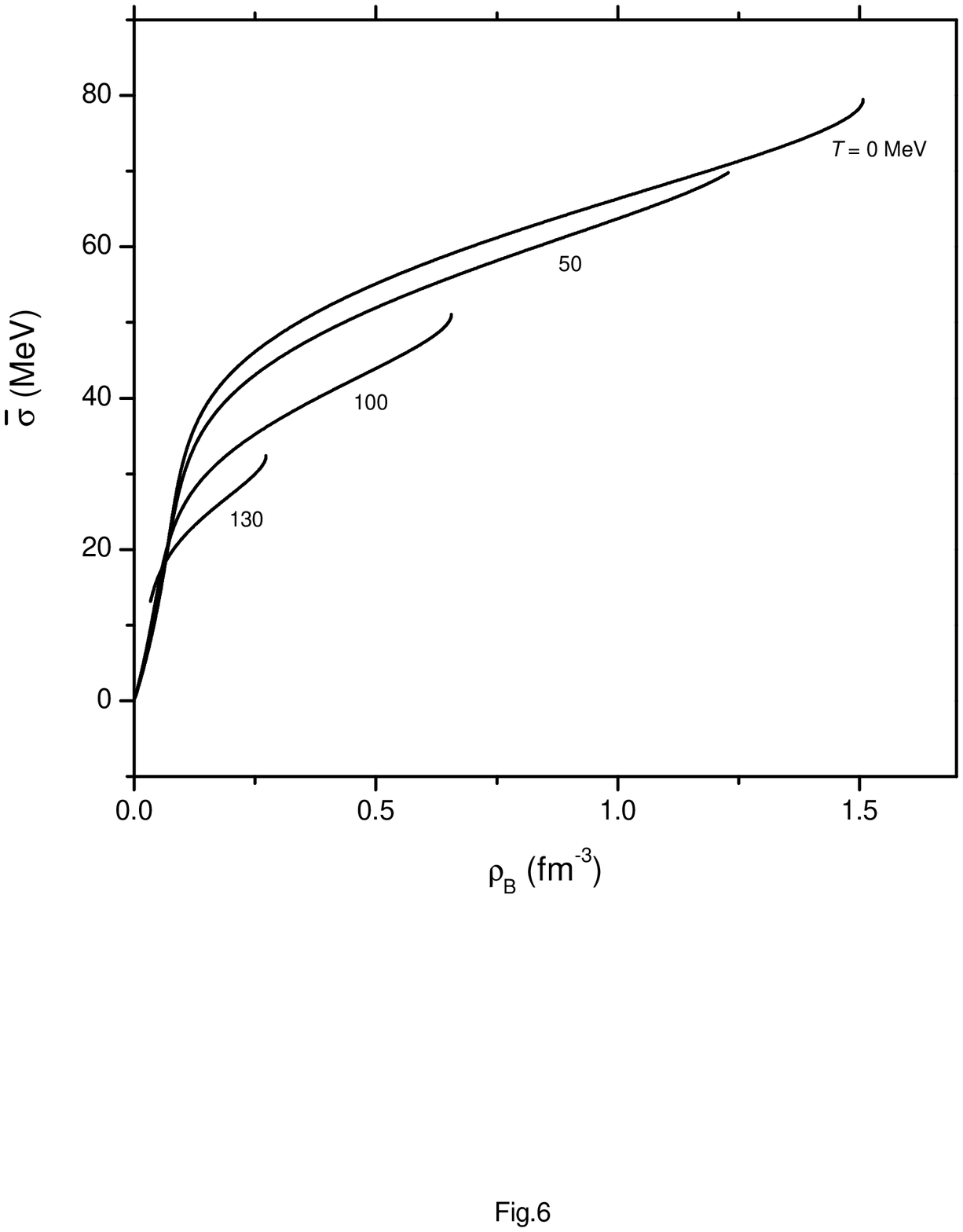}
\caption{$\bar{\sigma}$ vs. baryon density $\rho_B$ at different temperatures
$T = 0, 50, 100$ and $130$ MeV respectively}
\label{fig6}
\end{figure}

\begin{figure}[tbp]
\includegraphics[scale=0.8]{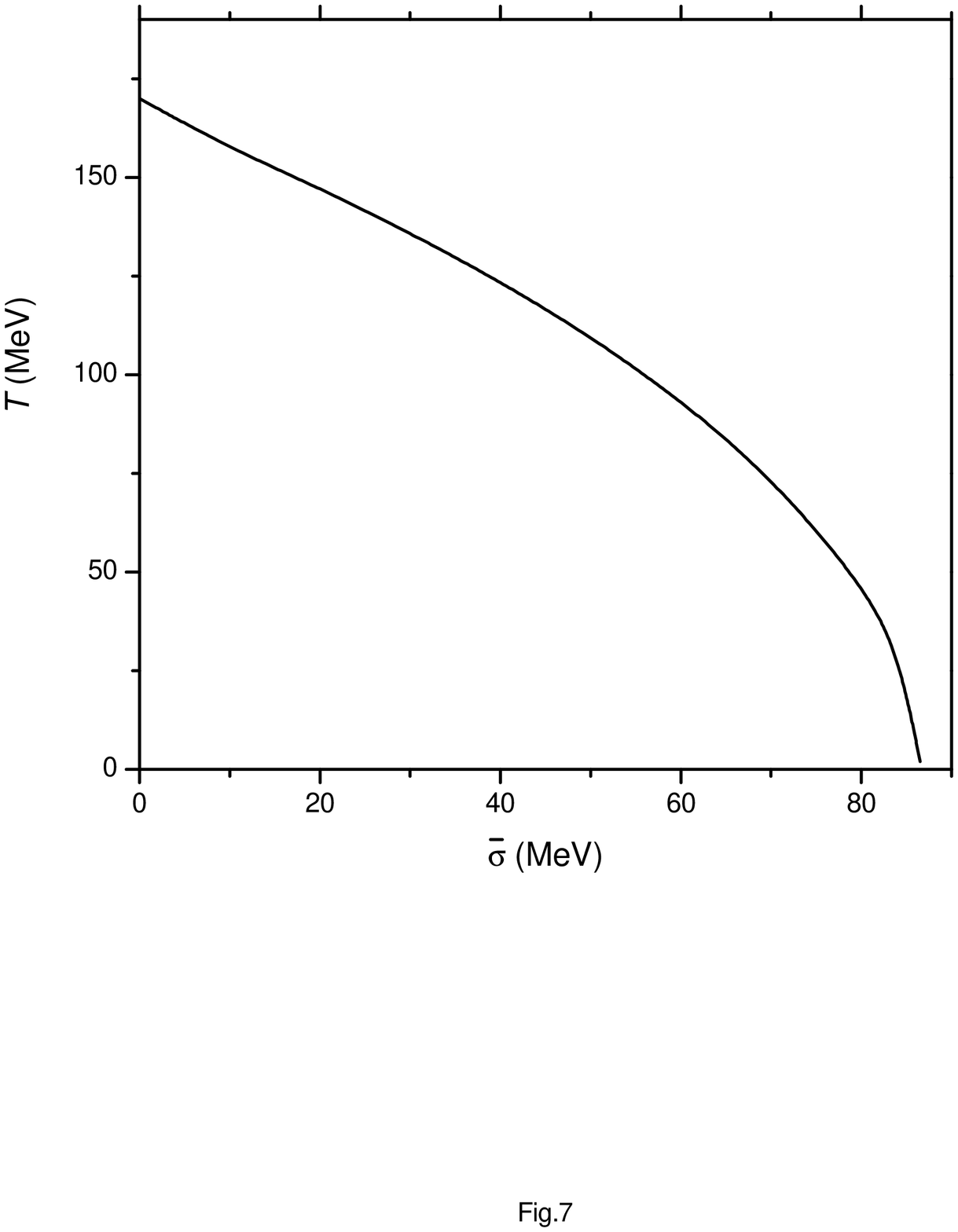}
\label{fig7}
\caption{Critical temperature of deconfinement transition $t_c$ vs. $\bar{\sigma}$}
\end{figure}

\begin{figure}[tbp]
\includegraphics[scale=0.8]{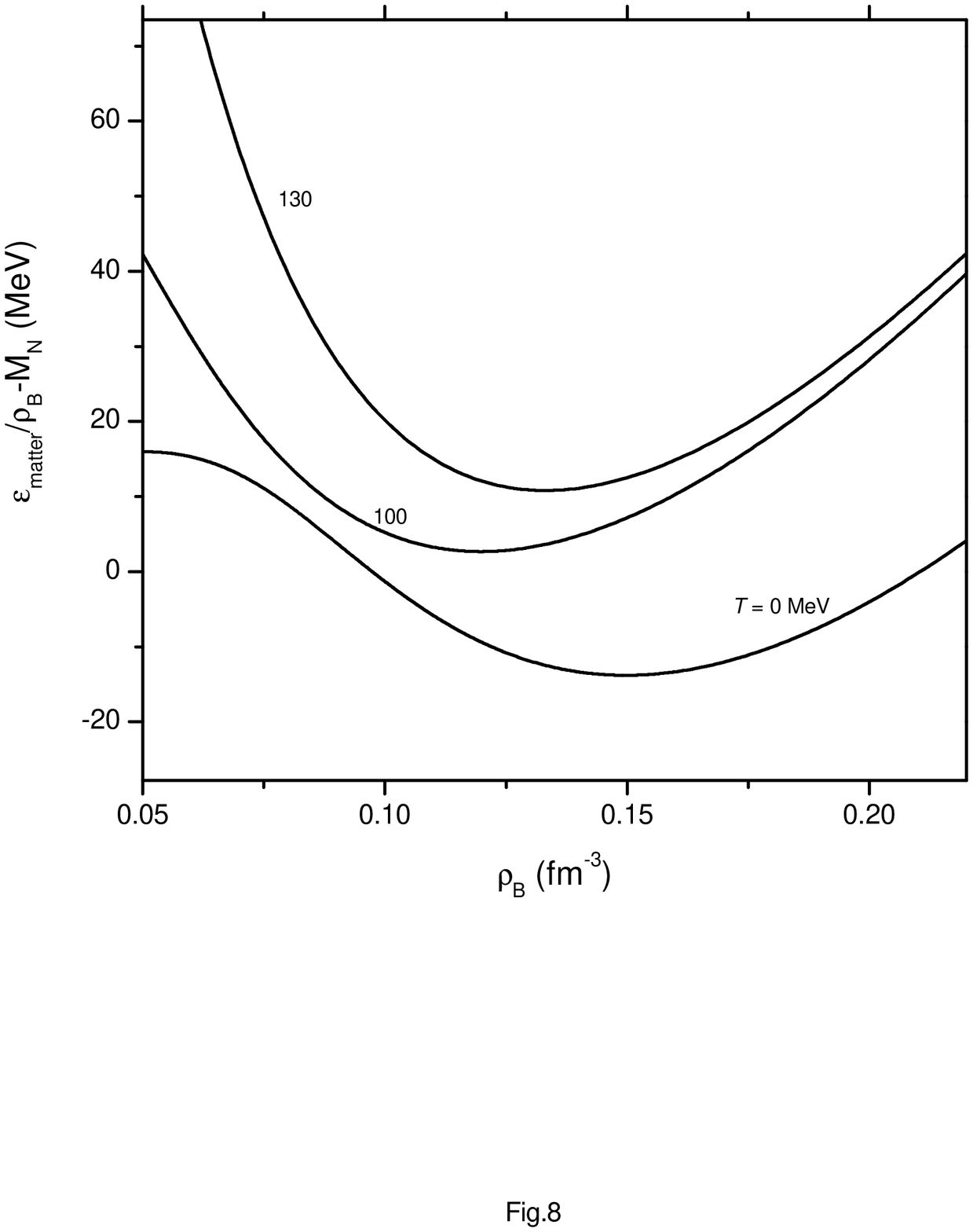}
\label{fig8}
\caption{Saturation curve of nuclear matter at different temperatures
$T = 0, 100$ and $130$ MeV respectively}
\end{figure}

\begin{figure}[tbp]
\includegraphics[scale=0.8]{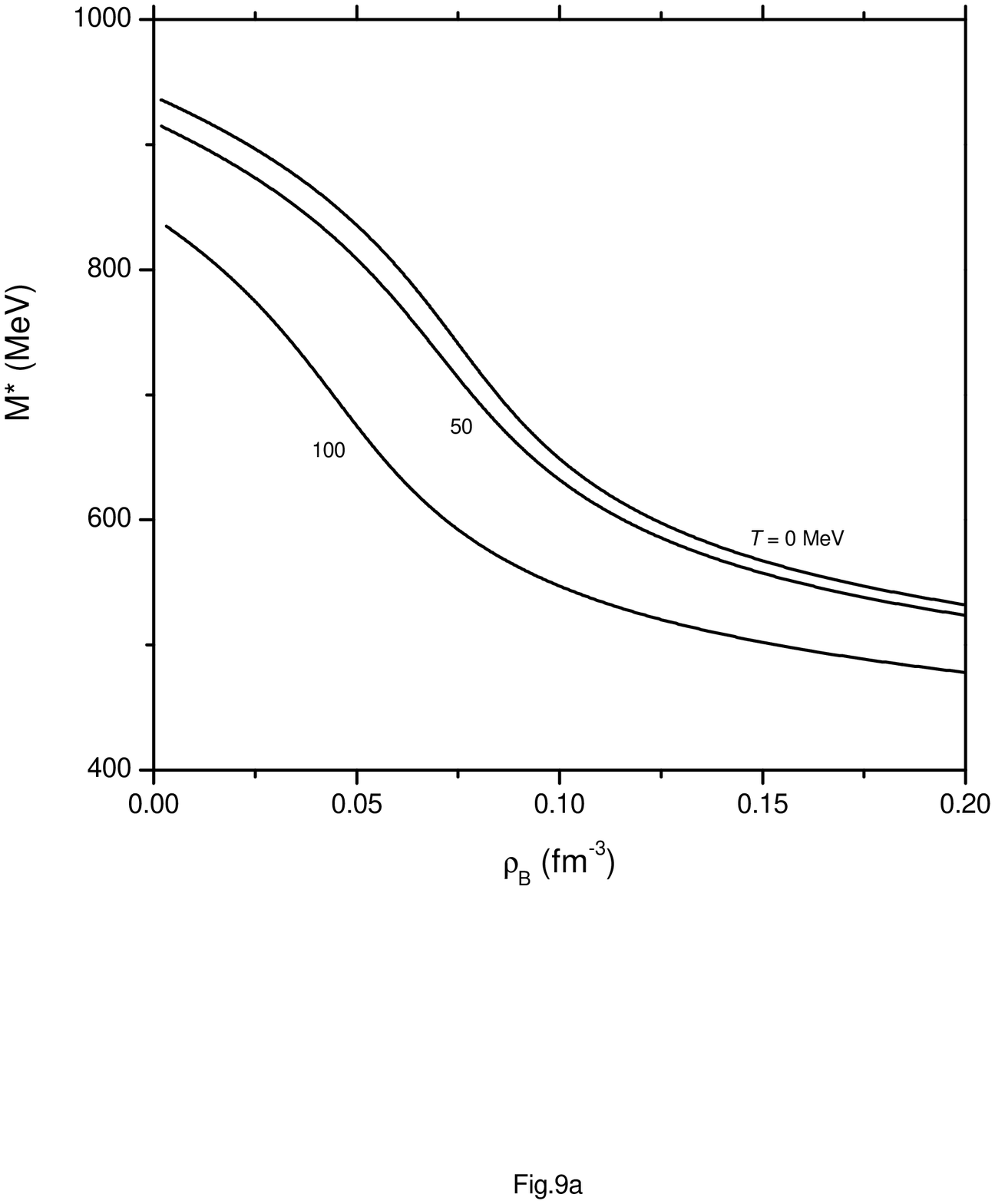}
\label{fig9a}
\caption{Effective nucleon mass vs. baryon density at different temperatures
$T = 0, 50$ and $100$ MeV respectively}
\end{figure}

\begin{figure}[tbp]
\includegraphics[scale=0.8]{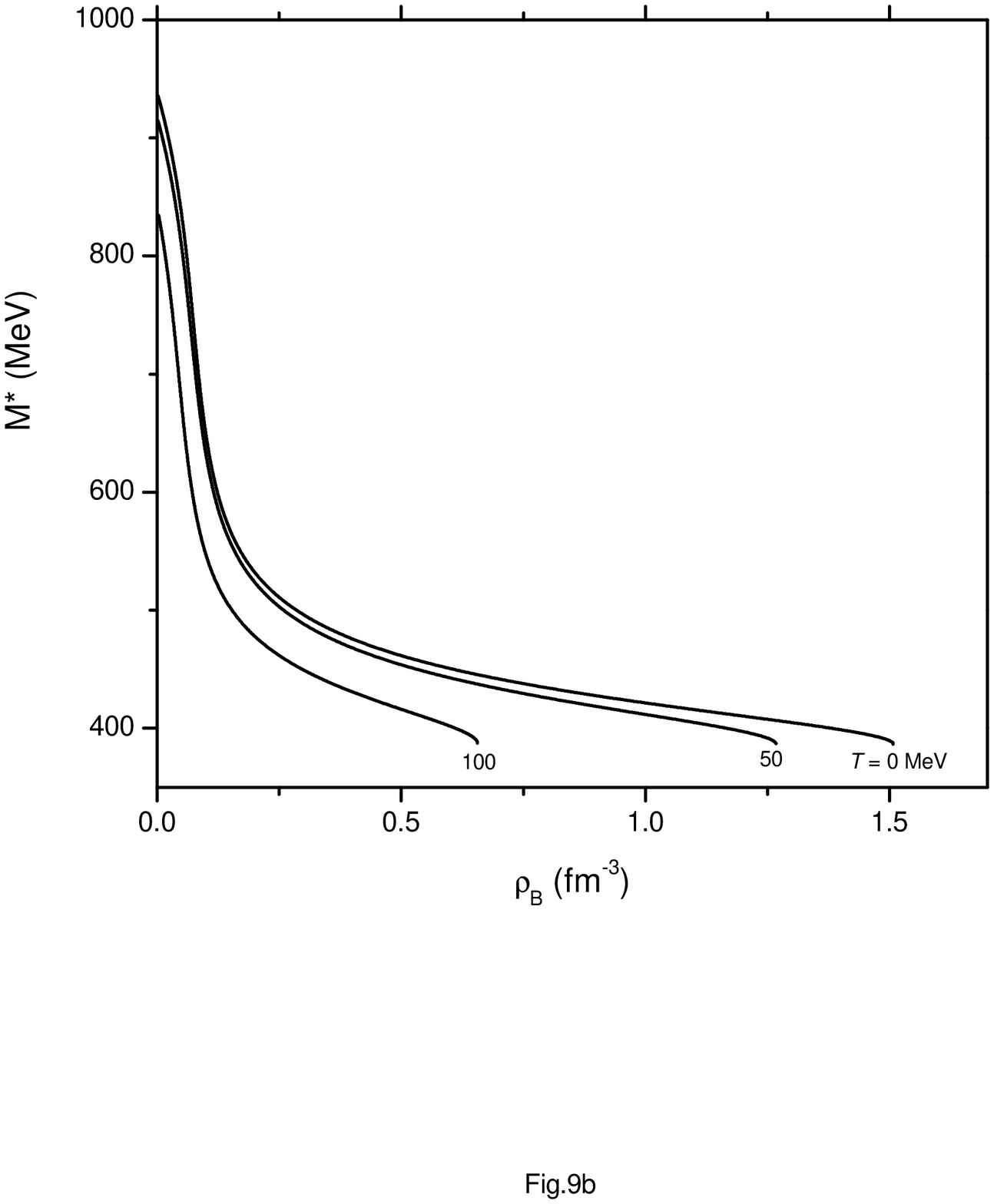}
\label{fig9b}
\caption{The same curve as Fig.4b with a wider range of baryon density at different temperatures
$T = 0, 50$ and $100$ MeV respectively}
\end{figure}

\end{document}